\documentclass[11pt,a4paper]{article}

\usepackage[utf8]{inputenc}
\usepackage[T1]{fontenc}
\usepackage{amsmath,amssymb}
\usepackage{booktabs}
\usepackage{graphicx}
\usepackage{hyperref}
\usepackage[margin=1in]{geometry}
\usepackage{enumitem}
\usepackage{xcolor}
\usepackage{multirow}
\usepackage{tabularx}

\newcolumntype{Y}{>{\raggedright\arraybackslash}X}

\hypersetup{
    colorlinks=true,
    linkcolor=blue,
    citecolor=blue,
    urlcolor=blue
}

\graphicspath{{../../implement/mirror/paper/figures/}}

\title{The Mirror Design Pattern: Strict Data Geometry over Model Scale\\for Prompt Injection Detection}

\IfFileExists{local_author.tex}{
  \author{J Alex Corll}

}{
  \author{
    The Parapet Project \\
    \url{https://github.com/Parapet-Tech/parapet}
  }
}

\date{}

\begin{document}

\maketitle

% ------------------------------------------------------------------
\begin{abstract}
% ------------------------------------------------------------------

Prompt injection defenses are often framed as semantic understanding
problems and delegated to increasingly large neural detectors. For the
first screening layer, however, the requirements are different: the
detector runs on every request and therefore must be fast,
deterministic, non-promptable, and auditable. We introduce Mirror, a
data-curation design pattern that organizes prompt injection corpora
into matched positive and negative cells so that a classifier learns
control-plane attack mechanics rather than incidental corpus shortcuts.
Using 5,000 strictly curated open-source samples---the largest corpus
supportable under our public-data validity contract---we define a
32-cell mirror topology, fill 31 of those cells with public data, train
a sparse character n-gram linear SVM, compile its weights into a static
Rust artifact, and obtain 95.97\% recall and 92.07\% F1 on a 524-case
holdout at sub-millisecond latency with no external model runtime
dependencies. On the same holdout, our next line of defense, a
22-million-parameter Prompt Guard~2 model reaches 44.35\% recall and
59.14\% F1 at 49\,ms median and 324\,ms p95 latency. Linear models
still leave residual semantic ambiguities such as use-versus-mention for
later pipeline layers, but within that scope our results show that for
L1 prompt injection screening, strict data geometry can matter more than
model scale.

\end{abstract}

% ------------------------------------------------------------------
\section{Introduction}
\label{sec:intro}
% ------------------------------------------------------------------

Prompt injection defenses are increasingly built around semantic models
that attempt to interpret whether a string of text is malicious. The
framing is understandable. Prompt injection is written in natural
language, so it is tempting to assume that the detector should also be a
language model. The problem is architectural. The first detector in the
path runs on every request. It becomes part of the security boundary. At
that point the detector needs properties that are not captured by a
benchmark score alone: low latency, stable behavior, transparent failure
modes, and resistance to prompt-based manipulation.

Ayub and Majumdar's CAMLIS 2024 result~\cite{ayub2024camlis} raised the
key question for this project. Their low-dimensional projections of
dense prompt embeddings did not suggest a clean linear boundary, which
made us ask whether prompt injection was intrinsically resistant to
simple classifiers or whether the difficulty was specific to the dense
semantic representations they studied. This paper starts from that
question: can a linear SVM distinguish benign requests from prompt
injection attempts at L1?

The L1 detector should minimize adaptivity, runtime complexity, and
prompt-sensitive behavior relative to the model it is protecting. It
should be static, discriminative, and easy to inspect. A promptable
model---one whose behavior can be altered by the content it is
classifying---in the hot path can expand the attack surface by inserting
another instruction-following system into the security boundary. That
may be acceptable for later layers that process a small residual, but it
is a poor default for a first-pass screen. In our implementation the L1
weights are compiled into the binary as a sparse perfect hash map, which
puts the detector in a different operational and attack-surface class
from a semantic model.

The main obstacle to this approach was not lack of model capacity. It
was lack of geometric discipline in the data. Public prompt injection
corpora are mixed in language, length, topic, formatting, and label
quality. If malicious and benign examples are not aligned across these
nuisance dimensions, a linear classifier learns shortcuts. It memorizes
corpus artifacts instead of injection structure. By \emph{geometry}, we
mean the intentional, cell-based pairing of contrastive malicious and
benign examples across nuisance dimensions such as language, length,
topic, and format, so that those variables stop defining the easiest
boundary. We call this pattern \emph{Mirror} because each malicious lane
is paired with a mirrored benign counterpart under the same cell
contract. Mirror was developed as a response to that problem and treats
curation as a geometric engineering task.

Our claim is narrow: strict data geometry can move the decision boundary
enough that a simple sparse classifier becomes a practical and effective
first-line defense for L1 prompt injection screening. The paper supports
that claim through three checkpoints: \texttt{v2}, which established the
initial Mirror baseline; \texttt{v3}, which showed that data geometry
materially changed performance with the model family held fixed; and
\texttt{v5}, which imposed a much stricter multilingual validity
contract and closed 31 of 32 cells with public data. Accordingly,
results are interpreted primarily within each checkpoint rather than as
a single directly comparable benchmark table across all versions.

\subsection{Task Definition and Scope}
\label{sec:task}

The evaluated L1 task is a binary prompt injection screening task, not a
general safety or jailbreak benchmark. Positive examples are
control-plane attacks: prompts or untrusted payloads that attempt to
override instructions, reassign roles, extract hidden system or tool
context, smuggle instructions through indirect data, or otherwise alter
the governing policy of the protected model. Operationally, Mirror
organizes this class into eight reason categories: instruction override,
roleplay jailbreak, meta-probe, exfiltration, adversarial suffix,
indirect injection, obfuscation, and constraint bypass.

Negative examples are benign requests, including instruction-following
tasks and security-adjacent text, that do not attempt to hijack the
model's control flow. We explicitly exclude pure content-safety
violations that may be unsafe but are not prompt injection---such as
toxicity, copyright, PII, and academic dishonesty---from the positive
class. Use-versus-mention ambiguity remains unresolved: quoted or
discussed attacks are a residual failure mode, as demonstrated in the
hard-benign challenge set (\S\ref{sec:challenge}).

Our contributions are:

\begin{enumerate}[nosep]
  \item We present Mirror as a reusable design pattern for prompt
    injection curation, based on matched positive and negative geometric
    cells.
  \item We show that prompt injection can admit a strong linear decision
    boundary for L1 screening when represented with sparse character
    n-grams and curated with strict geometry.
  \item We introduce a provenance-first curation and evaluation workflow
    that made leakage, source drift, and false occupancy visible instead
    of silently inflating results.
  \item We show that a 5k Mirror-trained linear SVM can substantially
    exceed the evaluated transformer baseline on the same held-out set
    while operating with far lower latency and far simpler deployment
    requirements.
  \item We characterize the residual failure modes---contextual
    ambiguity, use-versus-mention, and semantically thin attacks---that
    define the remaining architectural problem for replacing or reducing
    the L2a semantic layer.
\end{enumerate}

To keep those checkpoints explicit, Table~\ref{tab:checkpoints}
summarizes the versions discussed in the paper and the role each one
plays in the argument.

\begin{table}[t]
\centering
\caption{Checkpoint summary. Each row is a named dataset--model
configuration discussed in the paper.}
\label{tab:checkpoints}
\footnotesize
\setlength{\tabcolsep}{5pt}
\renewcommand{\arraystretch}{1.08}
\begin{tabularx}{\linewidth}{@{}p{2.6cm}p{3.5cm}Y@{}}
\toprule
Checkpoint & What changed & Role in argument \\
\midrule
\texttt{v2} \newline {\footnotesize 19k train / 2.4k holdout}
  & Initial Mirror baseline
  & Geometry baseline (0.835 F1); paired with \texttt{v3} to isolate the effect of data quality \\[6pt]
\texttt{v3} \newline {\footnotesize 19k locked / 1.9k holdout}
  & Source locking and provenance cleanup; model family held fixed
  & Geometry proof point: 0.918$\to$0.926 F1 from data discipline alone \\[6pt]
\texttt{v5} \newline {\footnotesize 5k curated / 524 holdout}
  & Strict multilingual validity contract; 31/32 cell closure
  & Headline result: 0.921 F1 / 0.960 recall; direct PG2 comparison \\
\bottomrule
\end{tabularx}
\end{table}

\paragraph{Reading guide.}
Section~\ref{sec:mirror} focuses on the geometry claim, with
\texttt{v2} included as the precursor needed to interpret the
\texttt{v2}$\to$\texttt{v3} change under roughly the same model family.
Section~\ref{sec:v5} turns to \texttt{v5}, asking whether the method
survives a much stricter validity and multilingual coverage contract.
The non-headline checkpoints are therefore included as architectural and
methodological landmarks, not as competing final benchmark claims.

% ------------------------------------------------------------------
\section{Related Work}
\label{sec:related}
% ------------------------------------------------------------------

Recent prompt injection detection work has generally assumed that better
semantic modeling should yield better detectors. Prompt Guard~2
(Meta, 2024)~\cite{pg2} is the most obvious practical reference point.
It is a compact transformer classifier (22M and 86M parameter variants)
designed for prompt injection and jailbreak detection, and it serves as
the primary baseline in this study because it represents the current
``small but semantic'' approach. This model is particularly relevant
here because it also appears as the next layer of defense (L2a) after
the linear classifier (L1) in the Parapet stack. Other public baselines,
including ProtectAI's DeBERTa-v3-based prompt injection
detector~\cite{protectai} (F1 0.539 on an earlier 2,386-case
\texttt{v2} holdout), make the same assumption: prompt injection is best
handled by a semantic model, even if the model is compressed or deployed
as a binary classifier rather than a generative LLM. ProtectAI's
training recipe was in fact the starting point for this project's own
data sourcing, before the Mirror pattern diverged the approach.

For the direct comparison in this paper, we evaluate the 22M-parameter
variant (the smallest and fastest PG2 checkpoint, making it the most
favorable baseline for a hot-path screening comparison), recorded in the
experiment manifests as \texttt{pg2-22m}.

Concurrent work such as PromptScreen~\cite{promptscreen} reaches a
similar architectural conclusion from a different direction.
PromptScreen uses a multi-stage pipeline with a TF-IDF linear SVM as
the core screening component and reports strong CPU-efficient
performance on a corpus of more than 30,000 prompts. That result
strengthens the argument that a simple linear detector belongs in the
hot path. The difference is that PromptScreen treats the model and
pipeline as the main story, while Mirror treats corpus geometry as the
main lever and asks why a linear screen can work so well with far less
data when the corpus is shaped correctly.

Our results point to a different representation-and-curation regime. We
move away from dense semantic embeddings and toward sparse character
n-grams, then constrain the corpus geometry so that language, format,
and related nuisance variables stop serving as cheap shortcuts. Under
that representation and data discipline, the task admits a much cleaner
linear boundary for L1 screening, with higher recall than the non-linear
models Ayub and Majumdar preferred in their setting.

More broadly, this is not unique to prompt injection. Prior
text-classification work has found that TF-IDF linear SVMs can remain
competitive with fine-tuned pre-trained language models on
domain-specific corpora, suggesting that larger semantic models are not
automatically the right default for every classification
setting~\cite{wahba2022}.

Mirror is adjacent to data-centric work on counterfactually augmented
data and spurious-correlation mitigation in
NLP~\cite{kaushik2020,kaushik2021,wang2022}. The difference is unit of
intervention and goal. Rather than rewriting individual examples into
counterfactual variants, Mirror treats the corpus itself as the unit of
design. Its aim is not only better classifier behavior, but an auditable
training contract: each cell has explicit scope, provenance, and pairing
logic.

Finally, this work sits within layered defense architectures rather than
replacing them. We do not argue against cascades. In fact, the system
design here depends on them. The claim is that the first layer should be
deterministic and cheap, while later layers can handle the smaller
residual set where semantic context genuinely matters.

% ------------------------------------------------------------------
\section{System Architecture}
\label{sec:arch}
% ------------------------------------------------------------------

This establishes the deployment context for the paper's claim: the
question is not only whether L1 scores well offline, but whether it can
function as a real first-pass gate inside a layered defense.

\begin{figure}[t]
\centering
\includegraphics[width=\columnwidth]{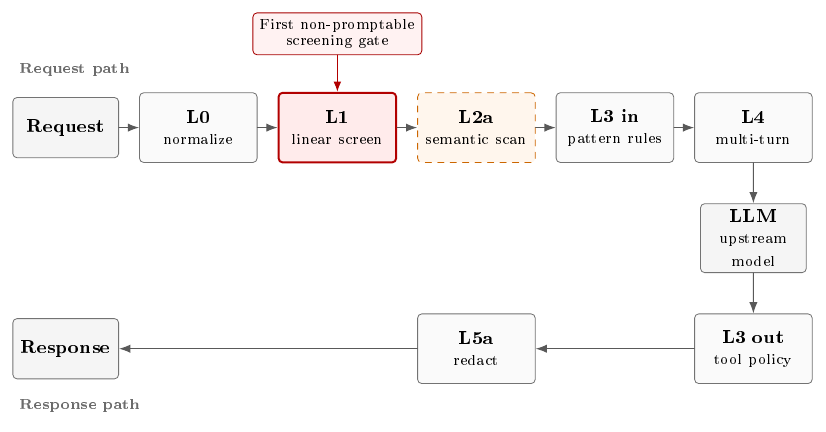}
\caption{The Parapet layered pipeline. L1 (this paper) applies a
compiled sparse linear screen on every request. Later layers handle the
residual.}
\label{fig:pipeline}
\end{figure}

Parapet began as a layered runtime defense rather than an isolated
classifier benchmark. The system established a request-path pipeline: L0
normalizes and assigns trust metadata, L1 applies a lightweight linear
screen (the focus of this paper), L2a runs an optional semantic scan for
residual ambiguity, L3 enforces deterministic pattern and policy rules,
and L4 adds multi-turn context before the request reaches the upstream
LLM. A symmetric response path handles output redaction and tool-call
policy. The practical goal was not to build a perfect detector in one
shot. It was to create a security boundary that could run on every
request without turning the detector itself into an operational or
attack-surface liability.

The L1 layer was simple from the start and remained simple in the final
system. Because the L1 architecture did not change across checkpoints,
we describe the \texttt{v5} form here. In its mature 5k form, training
uses \texttt{CountVectorizer} over sparse character word-boundary
n-grams with a \texttt{LinearSVC}, and the best run used $n=3..5$,
\texttt{min\_df=5}, binary presence, $C=1.0$, and no class weighting.
At runtime, the learned coefficients are compiled into the Rust binary
as a \texttt{phf} perfect hash map rather than shipped as an opaque
external model. Inference is a single bias-initialized dot product over
extracted n-grams, with the raw margin also passed through a sigmoid
when a probability-like score is needed for later combination. The
result is static, easy to inspect, and fast enough for always-on use.
There is no model server, no ONNX dependency, and no promptable
reasoning loop inside the screening layer. In the measured harness, the
L1 path runs in sub-millisecond time, while the Prompt Guard~2 baseline
operates at a median of 49\,ms and a p95 of 324\,ms on the same
machine.

The choice of character n-grams rather than word or subword features is
deliberate. A character n-gram window slides over the raw string
ignoring linguistic boundaries, which means it sees through common
evasion tactics that word tokenization cannot: spaced-out characters
(\texttt{s u d o}), Base64 fragments (\texttt{eyJ}, \texttt{==}), hex
encoding (\texttt{\%20}, \texttt{\textbackslash x0a}), and Unicode
substitution all leave contiguous character-level traces even when word
boundaries are destroyed. The tradeoff is vocabulary size---an
unconstrained character n-gram space over a large corpus can reach tens
of millions of features---which is why the production configuration caps
features at 15,000 and requires \texttt{min\_df=5}. Early specialist
experiments confirmed the choice empirically: word n-grams captured
roleplay and instruction-override patterns well but missed the
structural attack families (adversarial suffixes, obfuscation, indirect
injection) that character n-grams resolved.

This design was further informed by those specialist experiments. We
initially explored ensembles of category-specific models, each tuned for
a different attack family (e.g., word n-grams for roleplay jailbreak,
character n-grams without word boundaries for adversarial suffix). Those
experiments showed that the attack categories were real enough to
support specialized signal, but they also revealed the cost of runtime
specialization. Each additional specialist increased curation burden,
wiring complexity, and latency. That pushed the architecture back toward
a single strong generalist for L1. The specialist work did not
disappear, however. Its attack decomposition became the eight-category
reason taxonomy later used by Mirror: instruction override, roleplay
jailbreak, meta-probe, exfiltration, adversarial suffix, indirect
injection, obfuscation, and constraint bypass.

Mirror is a system to arrange data that enables building a first-pass
detector that fits a layered security system.

\begin{figure}[t]
\centering
\includegraphics[width=0.5\columnwidth]{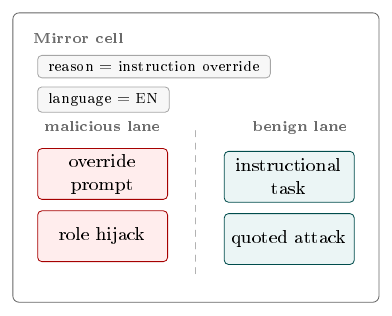}
\caption{From Mirror cell contract to compiled L1 artifact. Each cell
defines matched positive and negative examples; the SVM boundary is
compiled into a static Rust binary.}
\label{fig:mirror-to-l1}
\end{figure}

% ------------------------------------------------------------------
\section{The Mirror Design Pattern}
\label{sec:mirror}
% ------------------------------------------------------------------

This section isolates the geometry argument. Its role is to show,
through the \texttt{v2}$\to$\texttt{v3} transition, that changing corpus
structure while keeping the model family in place materially changed the
operating point.

\subsection{Why Geometry Matters}
\label{sec:why-geometry}

A linear classifier learns a boundary between positive and negative
examples. If the training distribution lets it separate those classes
using language, formatting, topic, or source artifacts, it will do that
instead of learning the structure of prompt injection. This is normal
empirical risk minimization on a badly shaped dataset.

Mirror architecture organizes the training corpus into cells of matched
positive and negative examples, forcing a linear classifier to learn the
boundary between them. In practice, that means aligning malicious and
benign examples across dimensions such as language, format, length, and
reason, so that those dimensions stop serving as spurious separators.
Mirror is not just balancing class counts. It is an attempt to control
the geometry of the training space.

This became the central hypothesis of the project after two earlier
observations. The first was leakage. Early results were inflated because
the data pipeline was not yet provenance-strict enough, and
train--evaluation contamination boosted performance. The second was
representational skepticism: if dense embedding spaces did not yield a
clean linear boundary, perhaps the obstacle was the representation and
corpus geometry rather than the task itself.

\subsection{Mirror Construction}
\label{sec:construction}

Mirror uses explicit cells as the unit of curation. A Mirror cell is a
curated subset of the training corpus indexed by attack reason and
language, containing matched positive and negative examples subject to
explicit admissibility and provenance rules. Matching aligns nuisance
dimensions---format, topic, approximate length---so that the remaining
discriminative signal is the control-plane attack pattern rather than a
corpus artifact. Examples that cannot honestly satisfy a cell's contract
are routed to residual or background pools rather than counted as
occupied.

In the \texttt{v5} checkpoint, cells are defined over eight attack
reasons and four languages for a 32-cell target topology. Each source is
annotated with scope and provenance rules.

This organization required software, not only labeling.
\texttt{parapet-data} provides the provenance pipeline and
\texttt{parapet-runner} the reproducible experiment
runner.\footnote{\url{https://github.com/Parapet-Tech/parapet}}
Together they track source hashes, semantic parity, manifests, and cell
fills. That infrastructure is not incidental. It is part of the method.
Without it, the project would have repeated the same leakage and
false-occupancy mistakes that partially motivated Mirror in the first
place.

\subsection{v2 to v3: Geometry as the Active Variable}
\label{sec:v2-v3}

The clearest evidence for Mirror came from the \texttt{v2} to
\texttt{v3} transition. With the same model family, the project moved
from a weaker geometry to a much stricter source-locked Mirror layout.
The historical \texttt{v2} holdout result was already competitive,
reaching 0.835 F1 with 0.931 recall on a 2,386-case holdout and
exceeding the transformer baselines on recall. But it still carried too
many false positives and too much residual impurity in the curation
logic.

The \texttt{v3} result was the real turning point. At the natural SVM
margin threshold of $t = 0.0$, the cleaned source-locked mirror baseline
reached 0.918 F1 (precision 0.932, recall 0.905, 62 FP, 90 FN on a
1,875-case holdout). After further cleaning, this improved to 0.926 F1
(precision 0.945, recall 0.908, 50 FP, 87 FN). The \texttt{v2} mirror
result on the same model family had been 0.835 F1 with 356 false
positives. The \texttt{v3} mirror cut false positives by more than 80\%
and gained 8--9 F1 points by reorganizing the same corpus into stricter
source-locked cells. The important point is not one decimal. It is the
direction of change under a fixed model family. Data geometry materially
changed the operating point.

The mixed-ratio ablation from \texttt{v2} supports the same
interpretation. These absolute F1 values reflect the \texttt{v2}-era
corpus and holdout; the important signal is the monotonic relationship,
not the scalar levels. When the training data was blended from pure
Mirror (100:0) to pure non-mirror (0:100) across three seeds, holdout F1
degraded monotonically (Table~\ref{tab:mix-ratio}).

\begin{table}[t]
\centering
\caption{Mixed-ratio ablation (\texttt{v2} corpus, three seeds).
Performance degrades monotonically as Mirror fraction decreases.}
\label{tab:mix-ratio}
\begin{tabular}{@{}lc@{}}
\toprule
Mirror : Non-mirror & Holdout F1 (mean $\pm$ std) \\
\midrule
100:0 & $0.837 \pm 0.005$ \\
70:30 & $0.825 \pm 0.004$ \\
50:50 & $0.819 \pm 0.008$ \\
30:70 & $0.819 \pm 0.003$ \\
0:100 & $0.788 \pm 0.002$ \\
\bottomrule
\end{tabular}
\end{table}

No sweet spot emerged from blending. Pure Mirror stayed best across all
seeds. This is exactly the pattern one would expect if geometry, not
just volume, were carrying the gain.

\begin{figure}[h]
\centering
\includegraphics[width=0.75\columnwidth]{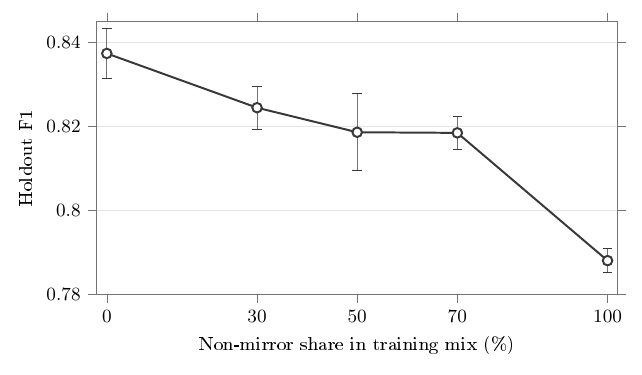}
\caption{Holdout F1 as a function of Mirror-to-non-mirror training
ratio (\texttt{v2} corpus, three seeds). Pure Mirror (100:0) is
consistently best; performance degrades monotonically.}
\label{fig:mix-ratio}
\end{figure}

\subsection{What the Model Learns}
\label{sec:features}

The feature story matters because it explains why Mirror is useful
rather than magical. Weakly structured baselines tended to learn keyword
fragments and substrings of ``secret,'' ``password,'' and
``confidential.'' Those features detect prompts that talk about
sensitive things, which is a proxy for the task but not the task itself.
Mirror-trained models were pushed toward injection mechanics instead:
instruction override cues, role and control-plane markers, adversarial
delimiters, and the structural traces of attempted hijacking. This is
still sparse lexical modeling, not semantics. But it is far closer to
the real task than a model that memorizes the vocabulary of
confidentiality.

The control-plane versus data-plane distinction is part of the method,
not an incidental cleanup pass. Mirror aggressively strips subjective
content-safety violations from the attack corpus so that L1 is trained
to police control-plane attacks---instruction hijacking and authority
transfer---rather than downstream data-plane categories such as
toxicity, copyright, or other policy misuse.

That design choice is one reason the resulting decision boundary is
cleaner. A classifier that is not asked to double as a general
content-safety filter can spend its capacity on the mechanics of prompt
takeover instead of on policy-adjacent lexical cues. In our
architecture, content safety sits outside the intended L1 scope: it is
broader, more subjective, and better handled by later semantic layers
than by the first-pass classifier.

The contamination story belongs here as well. Leakage was not a side
note; it changed the design of the project. The provenance pipeline
exposed 18.6\% train--evaluation contamination in the non-mirror
baseline and 7\% in the early mirror corpus. Without content hashing and
manifest tracking, the baseline would have appeared to crush the
evaluation because it had memorized nearly a fifth of the test set. The
fix was to carve the evaluation holdout first and assert zero train--eval
intersection via content hash before every run. Once those leaks were
repaired and the experiment rerun on clean splits, Mirror stopped being
just a clever sampling trick and became a stricter empirical contract.
That contract is what made later claims believable.

% ------------------------------------------------------------------
\section{Validity Recovery and Multilingual Closure: v5}
\label{sec:v5}
% ------------------------------------------------------------------

This checkpoint asks a different question from \texttt{v3}: not whether
Mirror can improve results under a fixed model family, but whether the
method still holds after much stricter validity, source-contract, and
multilingual coverage requirements are imposed.

Between \texttt{v3} and \texttt{v5}, error-driven corpus review showed
that content-safety contamination was a substantial component of the
source-pool defects. Across 167 adjudicated drops in the cleanup ledger,
34.1\% were explicit content-safety violations mislabeled as prompt
injection; this rises to 35.9\% when adjacent policy-not-PI categories
such as copyright, PII, and academic dishonesty are included. In the
stricter \texttt{v5\_r1} review tranche alone, explicit content-safety
cases accounted for 42.2\% of adjudicated drops. This motivated a
ground-up rebuild at lower volume with stricter label contracts.

That finding changed how we interpreted the earlier result. \texttt{v3}
had shown that mirror geometry mattered, but it had not yet imposed the
validity contract we ultimately wanted. For \texttt{v5}, we scaled down
to curate a smaller corpus we could defend cell by cell. In practice,
5,000 was not chosen because performance had clearly plateaued there; it
was the largest public-source corpus we could support while still
filling 31 of 32 cells under the stricter validity contract. By this
point it had become clear that occupancy was not enough. A cell with one
mislabeled row is not coverage. A multilingual pooled source is not
automatically valid mirror evidence for every reason--language pair it
touches. If the project was going to claim strict data geometry, it had
to stop hiding behind loose notions of coverage.

The \texttt{v5} recovery work formalized this. Coverage was redefined as
quota-based occupancy under explicit source contracts. The active
geometry was $8 \text{ reasons} \times 4 \text{ languages} = 32$
logical cells. Each cell needed reason-grounded malicious coverage and
valid benign support. Rows from pooled or mixed-scope sources could
still be useful, but they had to be routed as residual, background, or
quarantine unless they were promoted through a documented review path.

This made the data problem harder, not easier. The first \texttt{v5} run
covered only 18 of 32 cells; the final recovery line closed 31 of 32.
The recovery logic was staged rather than monolithic: broad passes were
used first to recover dense lanes and expose real deficits, then tiny
reviewed micro-sources and per-source tail extraction were used to close
sparse cells one by one. Chinese and Arabic were recovered through that
combination of broad sweeps and micro-sources. Russian proved harder and
shifted to tail-queue review across 444 rows from 10 source-specific
queues, which closed \texttt{indirect\_injection},
\texttt{exfiltration}, and \texttt{meta\_probe} but still left
\texttt{RU~constraint\_bypass} as the single accepted miss after direct
review found only 2 marginal candidates across the available sources
(0.5\% hit rate).

The Mirror method still worked after the evaluation contract was
tightened, the sources audited, and the gaps named rather than buried.

The final matrix is also useful as a compact demonstration of what
Mirror means in practice. It is not just abstract geometry. It is a
concrete multilingual corpus with explicit reason slots, explicit
failures, and explicit provenance. That is a stronger scientific object
than a single scalar F1.

\begin{figure}[t]
\centering
\includegraphics[width=0.65\columnwidth]{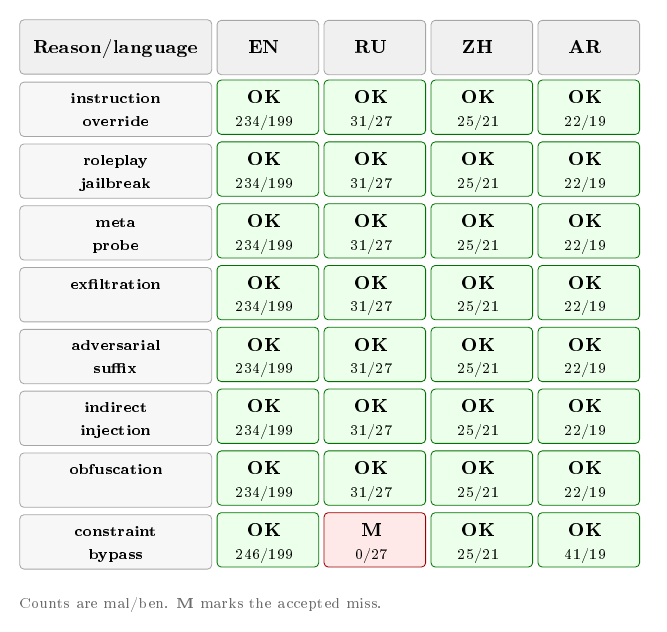}
\caption{Final \texttt{v5} coverage matrix: 8 reasons $\times$ 4
languages = 32 cells. Green = closed with valid data; red =
\texttt{RU~constraint\_bypass}, the single accepted miss.}
\label{fig:coverage}
\end{figure}

% ------------------------------------------------------------------
\section{Experimental Results}
\label{sec:results}
% ------------------------------------------------------------------

The evaluation contract for the final 5k comparison is summarized in
Table~\ref{tab:protocol}.

\begin{table}[t]
\centering
\caption{Evaluation protocol for the \texttt{v5} 5k comparison.}
\label{tab:protocol}
\small
\begin{tabular}{@{}p{3.8cm}p{8.5cm}@{}}
\toprule
Protocol item & Setting \\
\midrule
Holdout artifact & 524 cases total: 248 malicious, 276 benign \\
Split discipline & Holdout carved and frozen before training; training
  and evaluation are enforced content-hash disjoint \\
Evaluation preprocessing & L0 sanitization enabled at evaluation time
  for all compared detectors \\
Holdout composition & The 524-case holdout spans the same four-language,
  eight-reason task space as the training corpus, though not with
  uniform occupancy \\
Confidence intervals & 95\% bootstrap intervals with 10,000 resamples
  over the 524-case holdout \\
Threshold selection & L1 main results at fixed $t{=}0.0$ (natural SVM
  margin cutoff); PG2 evaluated at default thresholds
  (\texttt{pg\_threshold=0.5}, \texttt{block\_threshold=0.5}) \\
``Open-source'' in release terms & Training and evaluation examples are
  drawn from public-source corpora; the release publishes code, specs,
  manifests, and hashes needed to reconstruct the benchmark, not
  redistributed raw corpora or local runner artifacts \\
\bottomrule
\end{tabular}
\end{table}

\subsection{Main 5k Result}
\label{sec:main-result}

The current 5k control result comes from the final 5,000-sample corpus
built in the \texttt{v5} recovery line and a hyperparameter sweep over
the linear SVM family. The best reported run uses character
word-boundary n-grams with $n$ from 3 through 5,
\texttt{max\_features=15000}, \texttt{min\_df=5}, and $C=1.0$, and
reaches 0.9207 F1, 0.8848 precision, and 0.9597 recall on a 524-case
held-out evaluation set with 248 malicious and 276 benign examples. At
threshold 0.0 this corresponds to 31 false positives and 10 false
negatives. Bootstrap 95\% confidence intervals (10,000 resamples) are
$[0.895, 0.943]$ for F1, $[0.844, 0.921]$ for precision, and
$[0.934, 0.983]$ for recall. This is the clearest final demonstration
of the paper's empirical point: a small, static, carefully curated
linear model can become very strong when the data geometry is right.

The cryptographic content hashing and provenance trail matter here
because good numbers without provenance are not enough.

\subsection{Regex-Only Baseline}
\label{sec:regex}

The natural first question is whether machine learning is needed at all.
Our L3 regex layer---75 hand-written patterns covering 10 attack
categories---reaches 99.2\% precision but only 14.1\% recall (F1
24.7\%) on the same evaluation corpus. Regex is excellent at catching
known attack templates with near-zero false positives, but prompt
injection is too varied for static rules. The gap between 14\% recall
and 96\% recall is exactly what the learned L1 boundary provides.

\subsection{Semantic Baseline}
\label{sec:semantic}

The baseline protocol was designed to keep the comparison operationally
fair and mechanically simple. We evaluate \texttt{pg2-22m} off the
shelf, without task-specific fine-tuning, through the same
\texttt{parapet-eval} harness on the exact same 524-case holdout
artifact used for the SVM results, remapping the L2a decision to the
binary screening label. Both paths run with L0 sanitization enabled at
evaluation time. For PG2 we use the standard
\texttt{pg\_threshold=0.5} and \texttt{block\_threshold=0.5}
configuration; for L1 we report the compiled classifier at threshold
$t{=}0.0$, the natural margin cutoff used in the main holdout tables.
We choose the smaller \texttt{pg2-22m} variant because the comparison
target is hot-path screening, where latency is part of the result; we do
not present it as an upper bound on all possible semantic baselines. The
earlier ProtectAI DeBERTa baseline is not carried into the main table
because it was only evaluated in the historical \texttt{v2} protocol on
a different 2,386-case holdout, where it underperformed materially
(0.539 F1, 0.615 precision, 0.480 recall). After that result,
direct-comparison effort focused on Prompt Guard~2 as the stronger
semantic comparator. We also evaluated the 86M-parameter PG2 variant;
its recall profile was nearly identical to the 22M model at four times
the parameter count, confirming the smaller variant as the relevant
comparator.

\subsection{Head-to-Head Comparison}
\label{sec:h2h}

On the same 524-case holdout, the compiled L1 path ran at 0.32\,ms mean
latency (0.13\,ms median, 1.40\,ms p95) in the eval harness, with
earlier head-to-head runs landing around 0.26\,ms mean. Meanwhile the
\texttt{pg2-22m} Prompt Guard~2 baseline reached 0.5914 F1, 0.8871
precision, and 0.4435 recall, with 14 false positives and 138 false
negatives. The PG2 bootstrap intervals are $[0.529, 0.648]$ for F1,
$[0.828, 0.941]$ for precision, and $[0.381, 0.506]$ for recall. The
recall and F1 intervals are well separated, while the precision
intervals overlap. PG2 is conservative and precise on this set, but it
misses most malicious cases. The Mirror-trained L1 model gives up some
precision in exchange for a very large recall gain. For a first-pass
detector in a layered system, that trade is often correct.

The three approaches occupy different points on the same trade-off
surface (Table~\ref{tab:h2h}).

\begin{table}[h]
\centering
\caption{Head-to-head comparison on the 524-case \texttt{v5} holdout.}
\label{tab:h2h}
\begin{tabular}{@{}lcccccc@{}}
\toprule
Detector & Prec. & Recall & F1 & FP & FN & Latency \\
\midrule
L3 regex (75 patterns) & 0.992 & 0.141 & 0.247 & --- & --- & $<$1\,ms \\
Mirror L1 SVM (5k, $t{=}0.0$) & 0.885 & 0.960 & 0.921 & 31 & 10 & $<$1\,ms \\
Prompt Guard~2 (22M) & 0.887 & 0.444 & 0.591 & 14 & 138 & 49\,ms \\
\bottomrule
\end{tabular}
\end{table}

This comparison should be framed carefully. Once L1 has been rebuilt
around strict geometry, the semantic baseline no longer looks like the
obvious best answer for screening. It becomes a slower, narrower, and
more expensive signal. And the regex layer, while essential for catching
known templates at near-zero cost, cannot replace a learned boundary.
That is exactly why the open question is now about replacing or reducing
L2a rather than depending on it by default.

\subsection{Latency and Operational Profile}
\label{sec:latency}

The latency story is part of the result, not just an implementation
detail. In the measured harness, the L1 path operates in sub-millisecond
time, with historical measurements on the order of a few tenths of a
millisecond end to end. The PG2 baseline on the 5k holdout had a mean
latency of 109.2\,ms, a median latency of 49.4\,ms, and a p95 latency
of 324.4\,ms. Those numbers imply a different deployment model.

The same is true of the dependency profile. L1 is compiled into the Rust
binary as static weights. PG2 requires a model runtime and carries a
larger, more fragile systems footprint. Even if the raw classification
numbers were closer, this difference would still matter. Security layers
are operational systems, not just classifiers.

\begin{figure}[t]
\centering
\includegraphics[width=0.85\columnwidth]{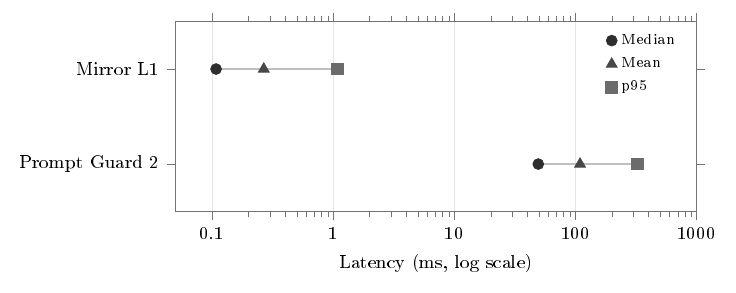}
\caption{Latency distributions for L1 (compiled SVM) vs.\ PG2 (22M).
L1 operates in sub-millisecond time; PG2 has a long tail reaching
324\,ms at p95.}
\label{fig:latency}
\end{figure}

\subsection{Reproducibility and Release}
\label{sec:repro}

The reproducibility story is part of the contribution. The curation
pipeline records source hashes, cell fills, gap reports, and semantic
hashes through \texttt{parapet-data}. The experiment runner records
training configuration, runtime hashes, evaluation output, and semantic
parity through \texttt{parapet-runner}. In practical terms, the paper is
supported by a toolchain that can rebuild named corpora from versioned
specs and rerun named experiments from recorded manifests and configs
rather than relying on hand-maintained notebooks or ad hoc scripts.
Compact specs expand into full cell-level specifications, making the
resulting mirror layout auditable rather than implicit.

The curation pipeline (\texttt{parapet-data}) and experiment runner
(\texttt{parapet-runner}) are released under the Apache~2.0 license at
\url{https://github.com/Parapet-Tech/parapet}. The \texttt{v5} results
reported in this paper correspond to the \texttt{mirror-paper-v5} tag.
Because training and evaluation data are drawn from third-party
public-source corpora with varying redistribution terms, the release
includes specs, manifests, and content hashes sufficient to reconstruct
the exact corpus and holdout from their original sources, rather than
redistributing the raw data directly.

\subsection{Challenge-Set Honesty}
\label{sec:challenge}

The model does have a ceiling. On an adversarial hard-benign challenge
set of 2,386 security-adjacent documents---whitepapers, CTF writeups,
red-team documentation that quotes or discusses attack patterns---the L1
SVM false-positive rate at threshold 0.0 was 51.9\%. Even at a raised
threshold of $+2.0$, the hard-benign FPR remained 12.3\%. On the same
challenge set, PG2 (22M) reached 21.3\% FPR---better, but still far too
high for production use. A 22-million-parameter semantic model cuts the
hard-benign FPR roughly in half relative to a sparse lexical model, but
it does not solve use-versus-mention disambiguation.
A separate challenge slice drawn from
JailbreakBench~\cite{jailbreakbench} paraphrase variants initially
appeared to expose a shared ceiling ($\sim$21\% recall for both the SVM
and Prompt Guard~2). Post-hoc inspection, however, showed that this
slice consists of direct harmful-content requests rather than prompt
injection: bomb-making, identity theft, hate speech, and similar
content-safety refusals, with no instruction override, role hijacking,
or indirect injection. We therefore no longer interpret that result as a
PI benchmark. Its only useful lesson is methodological: evaluation
slices must be scoped strictly to control-plane manipulation.

This is also the right bridge to the next design problem. The question
is no longer whether to replace the SVM with a bigger model. It is how
much of the current L2a residual can be replaced by a non-promptable or
at least non-generative second pass.

% ------------------------------------------------------------------
\section{Limitations}
\label{sec:limitations}
% ------------------------------------------------------------------

This paper is a proof of architecture and curation method, not a claim
that 5,000 samples are the final production recipe for prompt injection
detection. The project hit a real open-source data ceiling, especially
in multilingual attack coverage. The final \texttt{v5} matrix closed 31
of 32 cells, but it did so by careful sourcing and micro-source
construction, and one Russian lane remained an explicit accepted miss.
We therefore do not claim that performance plateaued at 5,000. It was
the largest corpus we could gather under the final public-data validity
contract. That is a success for methodological honesty, not evidence
that the data problem is solved.

The feature family also has clear limits. Character n-grams are very
good at capturing lexical and structural control-plane traces, but they
do not robustly resolve contextual ambiguity. A security whitepaper
quoting a jailbreak prompt can still look too much like an attack
(51.9\% FPR on the hard-benign challenge set at default threshold).
An earlier evaluation against JailbreakBench paraphrase variants
appeared to show both models plateauing at $\sim$21\% recall, but that
slice turned out to be content-safety material rather than prompt
injection (see Section~6.4). True paraphrase-robust PI evasion---attacks
that preserve injection semantics while eliminating lexical n-gram
traces---remains an open and likely architectural challenge for any
sparse linear model.

Although we report bootstrap intervals, the final holdout is modest in
size, and broader external validity across unseen production traffic
remains unproven.

The evaluation scope is also bounded by the datasets we were able to
build and audit. Public prompt injection datasets are noisy. Some mix
content safety with prompt injection. Some have poor provenance. Some
overrepresent English or a few familiar attack families. Mirror
mitigates this by making source contracts explicit, but it does not
magically produce a gold-standard universal benchmark.

Finally, the paper does not resolve the whole layered architecture. It
gives a stronger answer for L1 than we had before. It does not yet show
how to retire the current L2a semantic model without losing too much on
the residual set.

% ------------------------------------------------------------------
\section{Conclusion}
\label{sec:conclusion}
% ------------------------------------------------------------------

The main result of this paper is not that linear models beat
transformers in the abstract. It is that for L1 prompt injection
screening, strict data geometry can matter more than model scale. Once
the training corpus is organized as matched positive and negative cells,
with provenance and validity rules enforced as part of the method, a
simple sparse linear classifier becomes far more capable than the
standard story would predict.

The \texttt{v2} checkpoint established that this classifier might belong
inside a layered defense architecture. The \texttt{v3} checkpoint showed
that the geometry itself moved the decision boundary. The \texttt{v5}
checkpoint showed that the method could survive validity pressure,
multilingual recovery, and explicit source-contract accounting. The
final 5k result, 0.9207 F1 and 0.9597 recall on held-out data, is
therefore best understood as evidence for a design pattern.

The remaining problem is now clearer than it was at the start. Within
the scope of this study, the first layer is now empirically supported.
The open question is what should replace or reduce the current L2a
semantic layer. The honest answer is probably not ``nothing.'' L1 still
leaves a residual set defined by contextual ambiguity,
use-versus-mention cases, and semantically thin attacks. The next
research target is therefore a better residual architecture: one that
preserves the operational and security advantages of Mirror at L1 while
shrinking the need for a slow, promptable semantic detector at L2a.

If that problem can be solved, the generalization claim becomes much
stronger. But even without it, these results already suggest that
improving hot-path prompt injection defense may depend less on larger
hot-path models and more on better architecture, stricter geometry, and
semantic reasoning reserved for the cases that truly require it.

% ------------------------------------------------------------------
% References
% ------------------------------------------------------------------

\end{document}